\documentclass[letter]{aa}

\usepackage{graphicx}
\usepackage{txfonts}
\usepackage{ulem} 
\usepackage{amsmath}
\usepackage[]{hyperref}

\hypersetup{
    colorlinks = true,
    linkcolor = {blue}, 
    citecolor = {blue},
    urlcolor = {blue}
}

\usepackage{natbib}
\usepackage{color}
\usepackage{float}
\usepackage{xspace} 

\newcommand{\pipe}{\textsf{PIPE}\xspace}

\newcommand{\teff}{\ensuremath{T_{\mbox{\scriptsize eff}}}}

\begin{document}

\title{A CHEOPS White Dwarf Transit Search}

\author{Brett M.~Morris\inst{\ref{inst1}}
\and Kevin Heng\inst{\ref{inst1}}
\and Alexis Brandeker\inst{\ref{inst2}}
\and Andrew Swan\inst{\ref{inst3}}
\and Monika Lendl\inst{\ref{inst4}}}

\institute{Center for Space and Habitability, Gesellsschaftstrasse 6, 3012, Bern, Switzerland\label{inst1}
\and Department of Astronomy, Stockholm University, AlbaNova University Center, 10691 Stockholm, Sweden  \label{inst2}
\and Department of Physics and Astronomy, University College London, London WC1E 6BT, UK \label{inst3}
\and Observatoire de Gen\`eve, Universit\'e de Gen\`eve, Chemin des maillettes 51, 1290 Sauverny, Switzerland  \label{inst4}
}

  \date{Received 2021; accepted 2021}

  \abstract{White dwarf spectroscopy shows that nearly half of white dwarf atmospheres contain metals that must have been accreted from planetary material that survived the red giant phases of stellar evolution. We can use metal pollution in white dwarf atmospheres as flags, signalling recent accretion, in order to prioritize an efficient sample of white dwarfs to search for transiting material. We present a search for planetesimals orbiting six nearby white dwarfs with the CHEOPS spacecraft. The targets are relatively faint for CHEOPS, $11$ mag $< G < 12.8$ mag. We use aperture photometry data products from the CHEOPS mission as well as custom PSF photometry to search for periodic variations in flux due to transiting planetesimals. We detect no significant variations in flux that cannot be attributed to spacecraft systematics, despite reaching a photometric precision of $<2$ ppt in 60 s exposures on each target. We simulate observations to show that the small survey is sensitive primarily to Moon-sized transiting objects with periods $3$ hr $< P < 10$ hr, with radii $R \gtrsim 1000$ km. }
  
  
  \maketitle
  
\section{Introduction}

White dwarfs are the evolutionary end points of over 95\% of all stars \citep{Fontaine2001, Koester2013}. Given the ubiquity of small exoplanets orbiting Sun-like stars \citep{Petigura2018}, we seek to probe the following question: do rocky planets survive stellar evolution and continue orbiting their host remnants into the white dwarf phase?

The existence of planetesimals and dust grains orbiting main sequence stars has been firmly established by the discovery and study of debris disks \cite[for a review, see][]{Wyatt2008}.  Specifically, an infrared excess is detected relative to the stellar photosphere.  The lifetimes of the dust grains associated with this infrared excess are typically orders of magnitude shorter than the disk lifetimes, implying the existence of (invisible) planetesimals that are constantly replenishing the dust population via collisions \citep{Heng2010}.  An alternative interpretation of the infrared excess is that they are due to a large population of failed planetesimals with enough covering fraction to produce an observable infrared excess \citep{Heng2010}.  The study of debris disks motivates the study of observational signatures in white dwarfs that are of planetary origin.

There is clear spectroscopic evidence that rocky planetary material survives into the white dwarf phase \citep{Zuckerman1998, Zuckerman2003, Farihi2008, Zuckerman2007, Farihi2009, Farihi2010b, Farihi2010a, Farihi2012, Hermes2014, Rocchetto2015}. Spectroscopy reveals that nearly half of all white dwarfs show absorption lines due to metal pollution in their atmospheres \citep{Wilson2019}. Since metals have a finite lifetime in white dwarf atmospheres before sinking out of view, the presence of metals indicates recent accretion. Careful consideration of the chemical abundances present in white dwarf spectra have revealed that the accreted material must have been rock- and sometimes water-rich  \citep{Jura2009,Gansicke2012,Farihi2013}. Thus it appears that planetary bodies could survive into the white dwarf phase, which may provide a second era of habitability \citep{Agol2011}.

Infrared excesses at some WDs reveal the presence of debris discs containing $\sim$1000 K dust orbiting within 1 $R_\odot$, much smaller and warmer than their main-sequence counterparts \citep{Farihi2016}. These discs are found exclusively around polluted white dwarfs, and their infrared spectra have strong features from silicate grains, providing compelling evidence that the photospheric metals have a planetary origin \citep{vonHippel2007,Jura2007,Farihi2012}. The discs are hypothesised to be the end result of tidal disruptions of asteroids that have been perturbed onto highly eccentric orbits \citep{Jura2003}.

Photometry of WD 1145+017 revealed disintegrating, likely-rocky bodies orbiting a metal-polluted white dwarf \citep{Vanderburg2015}. Apparent transits of several dust clouds with orbital periods between 4 and 5 hr were observed by the {\it Kepler} space telescope. The transit depths measured as large as 40\%, and varied from transit to transit \citep{Izquierdo2018, Xu2018}. This system demonstrates the great potential of metal-polluted white dwarf planetary systems: (1) the spectroscopic detection of metal pollution enhances transit search efficiency by looking at only white dwarfs with recent accretion histories; (2) the geometrical advantage afforded by the small radii of white dwarfs produces deep transits from small objects; (3) the orbits where transiting material may be found have periods of only a few hours; and (4) each transit transmits unique information about the rapidly changing system, which appears to evolve from one orbit to the next. 

Since the 2015 discovery, several further detections of transiting material have been made including the discovery of 20 day-long eclipses with a period of 107 days in optical photometry of ZTF J013906.17+524536.89 \citep{Vanderbosch2020}, and the discovery of transits of WD 1856+534 with a period of 1.4 days, corresponding to a Jupiter-sized occultor \citep{Vanderburg2020}.

The CHEOPS mission is a 30 cm effective-aperture telescope in low-Earth orbit since 2019 \citep{Deline2020, Futyan2020, Hoyer2020}. The spacecraft design is specified in \citet{Benz2020}, and its initial performance is characterized by \citet{Lendl2020}. In the first call for CHEOPS Guest Observer programs, we were awarded 84 orbits to search for transiting debris orbiting white dwarfs. We present the set of observations in AO-1 Program 7, completed in February 2021. 

In Section~\ref{sec:theory}, we provide a brief theoretical motivation for searches for transiting material on short orbital periods. In Section~\ref{sec:obs}, we share two parallel analyses of the CHEOPS photometry with the data products produced by the CHEOPS team, as well as a reduction by a custom PSF photometry package. In Section~\ref{sec:sim}, we summarize a small simulation to estimate the detection efficiency of our CHEOPS observations in terms of sensitivity to planetsimal orbital period and radius. We discuss the implications of our observations in Section~\ref{sec:discussion} and conclude in Section~\ref{sec:conclusion}.

\section{Theoretical motivation} \label{sec:theory}

\begin{figure*}
    \centering
    \includegraphics{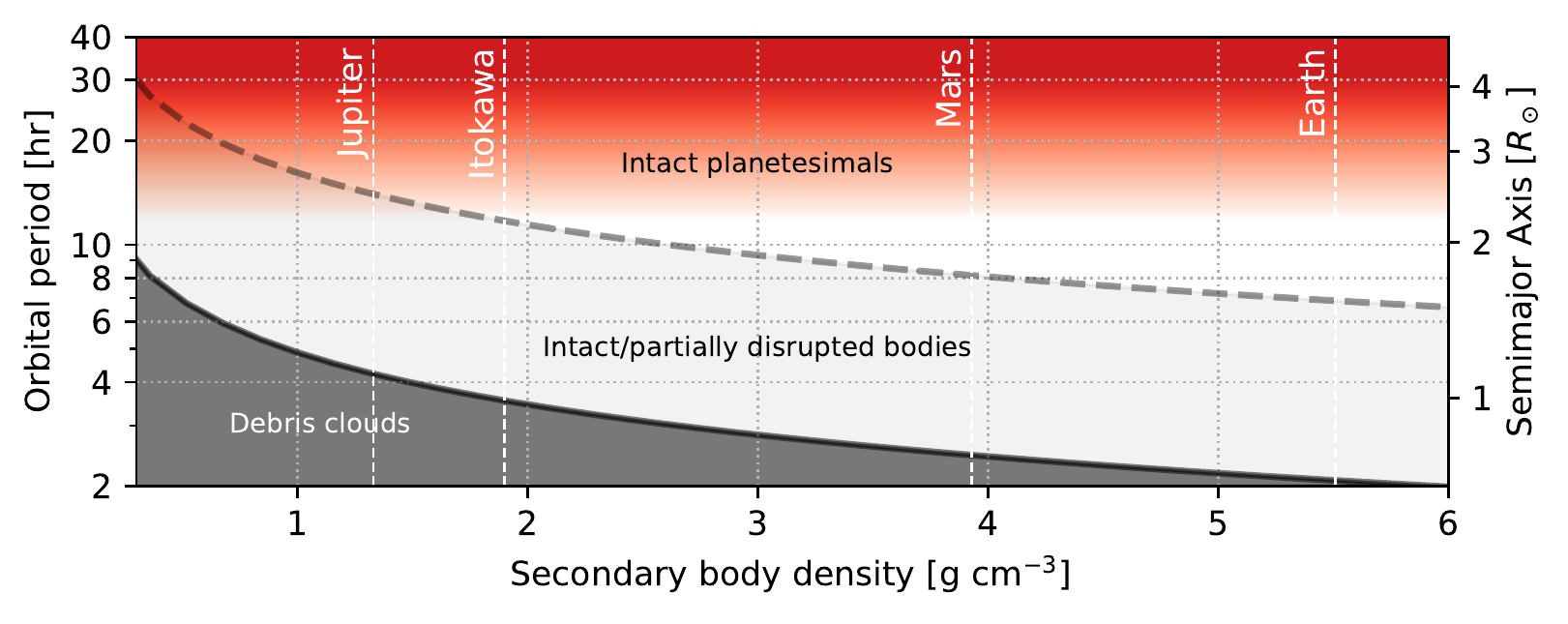}
    \caption{Tidal disruption limits in period-density space for circular orbits \citep{Veras2014}. Vertical dashed lines mark the densities of several secondary bodies spanning a factor of five in bulk density from gas giant planets and rubble-pile asteroids through dense rocky bodies. Red regions of the plot represent regions where only single transit events may be detected in the CHEOPS photometry, non-red regions note where several occultations are possible within 24 hours of photometry. The curves denote the regions where intact bodies may reside, the middle region marks potentially disrupted bodies, and black denotes bodies that we expect to be tidally disrupted. }
    \label{fig:disruption}
\end{figure*}

The tidal disruption radius $r_c$ for circular orbits is approximately
\begin{equation}
    \frac{r_c}{R_\odot} = C \left(\frac{M_\mathrm{WD}}{0.6 \mathrm{~M}_\odot}\right)^{1/3} \left(\frac{\rho}{3\mathrm{~g~cm}^{-3}}\right)^{-1/3},
\end{equation}
where $C$ is a constant ranging from about 0.85 to 1.89 \citep[see e.g.:][]{Veras2014,Bear2013}. The period-density space of disrupted and intact bodies is shown in Figure~\ref{fig:disruption}. 

As we will discuss, we have observed the sample of white dwarfs in Table~\ref{tab:targets} for 24 hours each, modulo the window function of Earth occultations. We required the minimum number of detected transits to be at least two in order to measure a periodic transiting event, so we only consider orbital periods $\lesssim 12$ hours. With these observations in mind as a guide, Figure~\ref{fig:disruption} suggests that rocky planetesimals with bulk densities like most small asteroids, $\rho \approx 1.9$ g cm$^{-3}$ \citep{Fujiwara2006}, should be tidally disrupted on orbital periods $P \lesssim 4$ hours, may be tidally disrupted on $4$ hr $< P < 12$ hr, and likely are intact otherwise. 

By surveying white dwarfs for objects with orbital periods $<12$ hours, we are primarily sensitive to tidally disrupted material that has nearly circularised. This is consistent with the expectation that there is rocky material falling onto these white dwarfs based on the spectroscopic observations of metal pollution in their atmospheres.

\section{Observations} \label{sec:obs}

The sample of metal-polluted white dwarfs observed with CHEOPS in AO-1 Program 7 is enumerated in Table~\ref{tab:targets}. It includes six white dwarfs closer than 25 pc which span a factor of four in \teff.

In this work we consider photometry produced with two methods, each processed with the same linear detrending as outlined in Section~\ref{sec:linea}; in Section~\ref{sec:drp} we describe the standard aperture photometry products, and we revisit the same observations in Section~\ref{sec:psf} with point-spread function (PSF) fitting photometry.

\subsection{Linear detrending} \label{sec:linea}

The fluxes measured by the default aperture photometry from CHEOPS, $f$, contain several systematic trends as a function of roll angle. To first order, the flux that we observe is a linear combination of the astrophysical signals we wish to detect and some unknown function of observational basis vectors. For some CHEOPS programs, the astrophysical signal has unconstrained quantities, like transit times or depths. In principle, there is some design matrix $\bf X$, which we can solve for the least-squares estimators $\hat{\beta}$ such that
\begin{equation}
{\bf X} \hat{\beta} = f.
\end{equation}
We provide a Python package for assembling a suitable design matrix from the CHEOPS data products called \textsf{linea}\footnote{\url{https://github.com/bmorris3/linea}}, which by default concatenates a design matrix of a unit vector, and the cosine and sine of the roll angle, and we add the product of the sine and cosine of the roll angle as an additional basis vector. The roll angle terms account for the variations in flux that occur as a result of the rotation of the spacecraft.

\subsection{CHEOPS Data Reduction Pipeline (DRP) Photometry} \label{sec:drp}

The DRP photometry contains corrections for dark and flat field normalization, bad pixels (pixels that are either hot or dead, and cosmic ray hits), and contamination from nearby stars in the aperture \citep{Hoyer2020}. After these corrections, the DRP measures the aperture photometry in four apertures of varying radii -- we find that the DEFAULT aperture photometry produces a small median absolute deviation in the finished light curve, so we rely on that reduction for the DRP results presented here. The aperture radii for each target are listed in Table~\ref{tab:targets}.

We mask outliers in the time series photometry by their corresponding stellar centroids by measuring the $4\sigma$ interval over which most stellar centroids reside, and rejecting any fluxes for which the centroid exceeds this radius from the mean centroid position. On average, we reject 4\% of fluxes with this approach.

\begin{figure*}
    \centering
    \includegraphics[width=\textwidth]{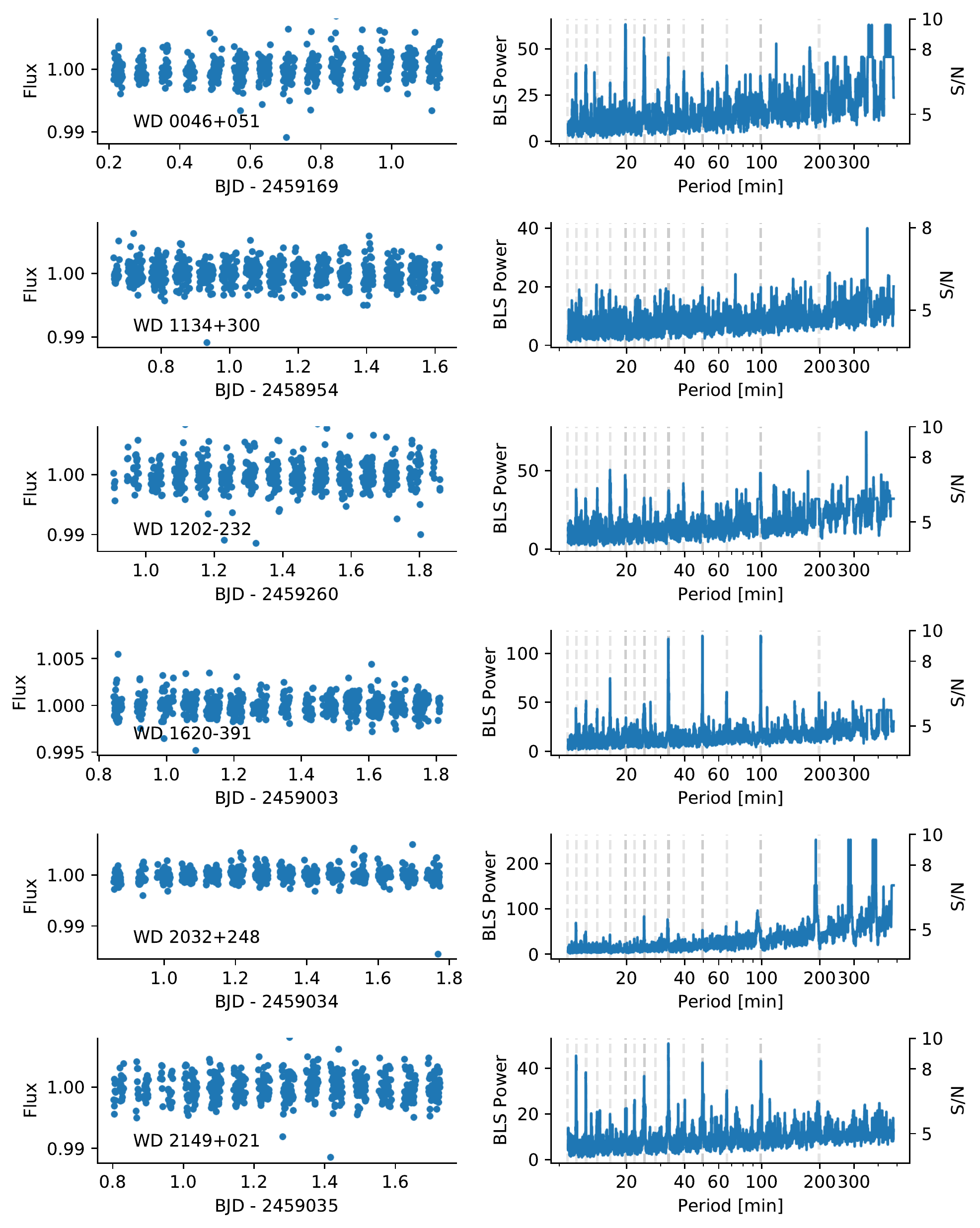}
    \caption{CHEOPS Data Reduction Pipeline (DRP) light curves of six white dwarfs observed with the CHEOPS spacecraft in 2020 and 2021. Dashed vertical gray lines represent aliases of the CHEOPS orbital period ($\sim100$ minutes). }
    \label{fig:gallery}
\end{figure*}

The DRP photometry reduced with the linear detrending described in Section~\ref{sec:linea} is shown in the left panels of Figure~\ref{fig:gallery}, and the corresponding box-least squares (BLS) periodograms are shown in the panels on the right. The DRP photometry has typical median absolute deviations similar to 1 ppt in 1 min exposures. Periodic signals are detected in the photometry that can usually be ascribed to aliases of the CHEOPS orbital period of $100$ minutes (marked with vertical dashed lines in the periodograms). We observed that periodic single-point outliers tend to occur at the beginning and end of an orbit which correspond to the times when the Earth is closest to the CHEOPS field of view, when stray light can be expected to be greatest (see Appendix~\ref{app:b}).

\subsection{PSF Photometry} \label{sec:psf}

To get an independent photometric extraction and potentially improve the precision, we adopted the PSF extraction tool \textit{PSF Imagette Photometric Extraction} (\pipe) to use on the subarrays. Some advantages of using PSF photometry compared to aperture photometry are that: (1) it can properly weigh the contribution to the signal of each pixel within the aperture to reduce the impact of the readout noise; (2) it is easier to filter out the impact of hot pixels and cosmic rays by identifying outlier pixels and removing them from the fit; and (3) the background can be fit simultaneously with the PSF to be less sensitive to spatial variations. A potential problem with this technique is that it can introduce noise due to a mismatch between the model and the actual PSF. CHEOPS was built to have a stable PSF, but in particular jitter (typically $< 1$\,pix) will introduce motion blur during the 60\,s exposures and modify the measured PSF. \pipe deals with motion blur by fitting a linear combination of model PSFs slightly offset from each other to the imagettes, improving the fit and leaving residuals dominated by photon noise. The model PSFs themselves were derived, one for each visit, by combining all observed frames within that visit.

\begin{figure*}
    \centering
    \includegraphics[width=\textwidth]{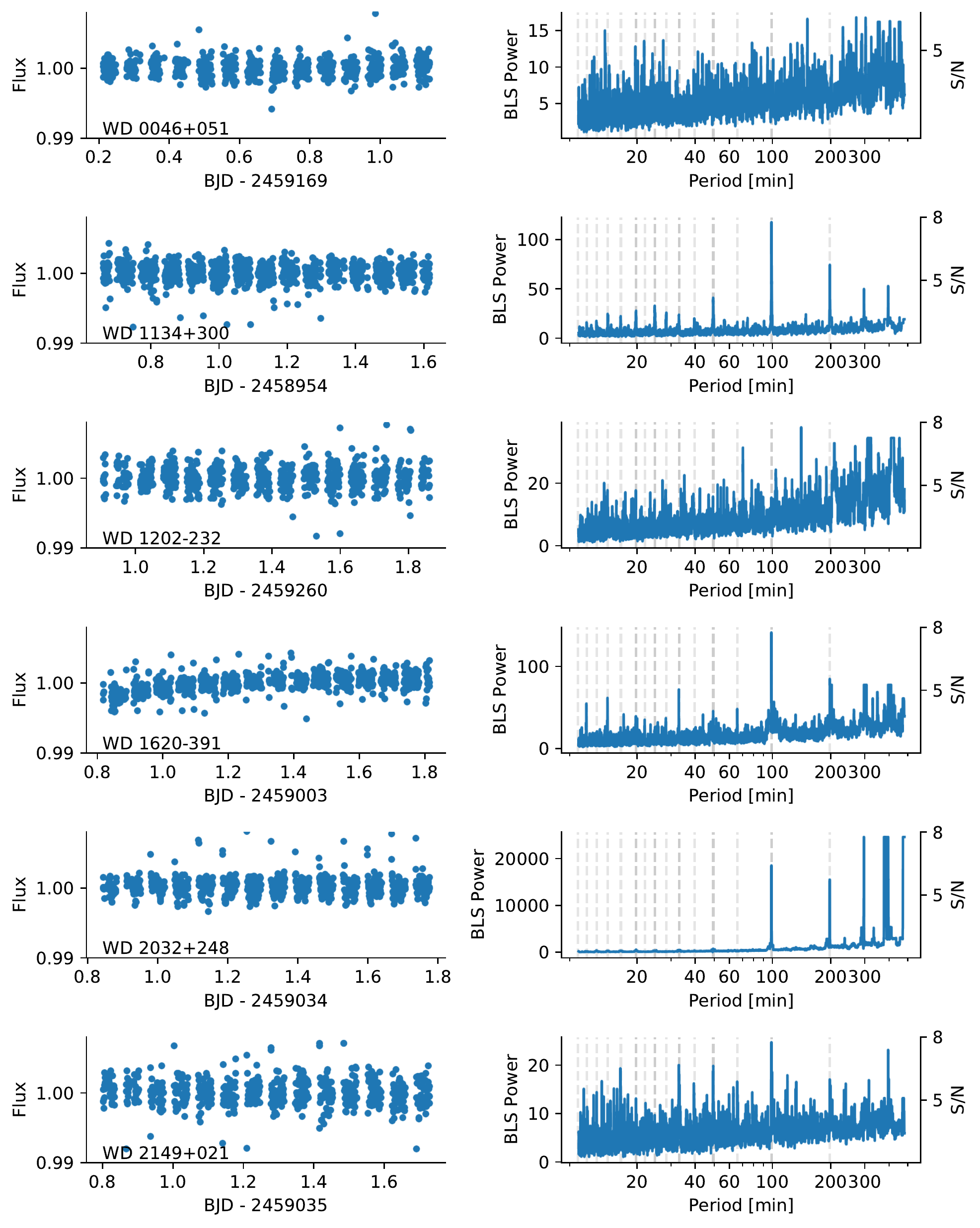}
    \caption{PIPE PSF photometry Light curves of six white dwarfs observed with the CHEOPS spacecraft in 2020 and 2021. Compared with the DRP photometry in Figure~\ref{fig:gallery}, the \pipe photometry has a different sensitivity to detector artifacts and outliers.}
    \label{fig:gallery_pipe}
\end{figure*}

The \pipe photometry, also reduced with the linear technique in Section~\ref{sec:linea}, is shown in Figure~\ref{fig:gallery_pipe}. The \pipe photometry has typical median absolute deviations similar to 1 ppt in one-minute exposures, like the DRP photometry, but performs better by up to a factor 2 for the fainter targets. As with the DRP photometry, periodic signals are detected in the photometry which are usually aliases of the CHEOPS orbital period (marked with vertical dashed lines).

\begin{table*}
\centering
\begin{tabular}{lccccrccc}
WD & Date & Aperture  &Dist. & Gaia $G$ & \teff$^{a}$ & Spectral & DRP MAD & PIPE MAD \\
   & [UTC] & Radius [pix] &[pc] & [mag] & [K] & Type  & [ppt] & [ppt] \\\hline \hline
0046+051 & 2020-11-15 & 25 & 4.3   & 12.31 & 6,400  & DZ$^b$  & 1.65 & 1.25\\
1134+300 & 2020-04-15 & 33 & 15.7  & 12.51 & 22,000 & DA$^c$ & 1.59 & 1.48 \\
1202--232 & 2021-02-15 & 25 & 10.4  & 12.75 & 8,700  & DAZ$^d$ & 1.98 & 0.98\\	
1620--391 & 2020-06-03 & 25 & 12.9  & 11.00 & 26,200 & DAZ$^e$   & 1.00 & 1.07 \\
2032+248 & 2020-07-04 & 25 & 14.8  & 11.55 & 20,400 & DAZ$^f$ & 1.03 & 1.66\\
2149+021 & 2020-07-05 & 25 & 22.5  & 12.79 & 18,100 & DAZ$^g$ & 2.13 & 1.14\\ \hline
\end{tabular}
\caption{CHEOPS white dwarf survey sample and the light curve median absolute deviation precision in one minute exposures in parts per thousand for the CHEOPS Data Reduction Pipeline (DRP) and the \pipe PSF photometry. (a) Temperatures compiled by \citet{GentileFusillo2019}; spectral types from (b) \citet{vanMaanen1917}, (c) \citet{McCleery2020}, (d) \citet{Zuckerman2003}, (e) \citet{Holberg1995}, (f) \citet{Wesemael1984}, (g) \citet{Koester2005}. \label{tab:targets} }
\end{table*}

\section{Transit Simulations} \label{sec:sim}

We determine the probability of detecting a single transiting object with time series photometry by generating a series of $10^6$ transit models, and injecting them into the observations of our brightest and faintest targets using the open source Python package \textsf{aspros}\footnote{\url{https://github.com/bmorris3/aspros}}. We inject transits with the \citet{Mandel2002} transit light curve model with orbital periods ranging from $3$ hr $< P < 12$ hr, transiting object radii $500$ km $< R < 2000$ km, and assuming a fiducial white dwarf with $R_\mathrm{WD}=9\times10^8$ cm and $M_\mathrm{WD}=0.6 M_\odot$. 

We compute a BLS periodogram for each light curve, and measure the period with the strongest power and its aliases \citep{Kovacs2002}. We measure the approximate signal-to-noise (S/N) of the candidate transit observations by measuring the significance of a transit model (fixed depth) compared with a transit-less model. If the $S/N > 10$, we consider the transit detected. 

\begin{figure}
    \centering
    \includegraphics[scale=0.8]{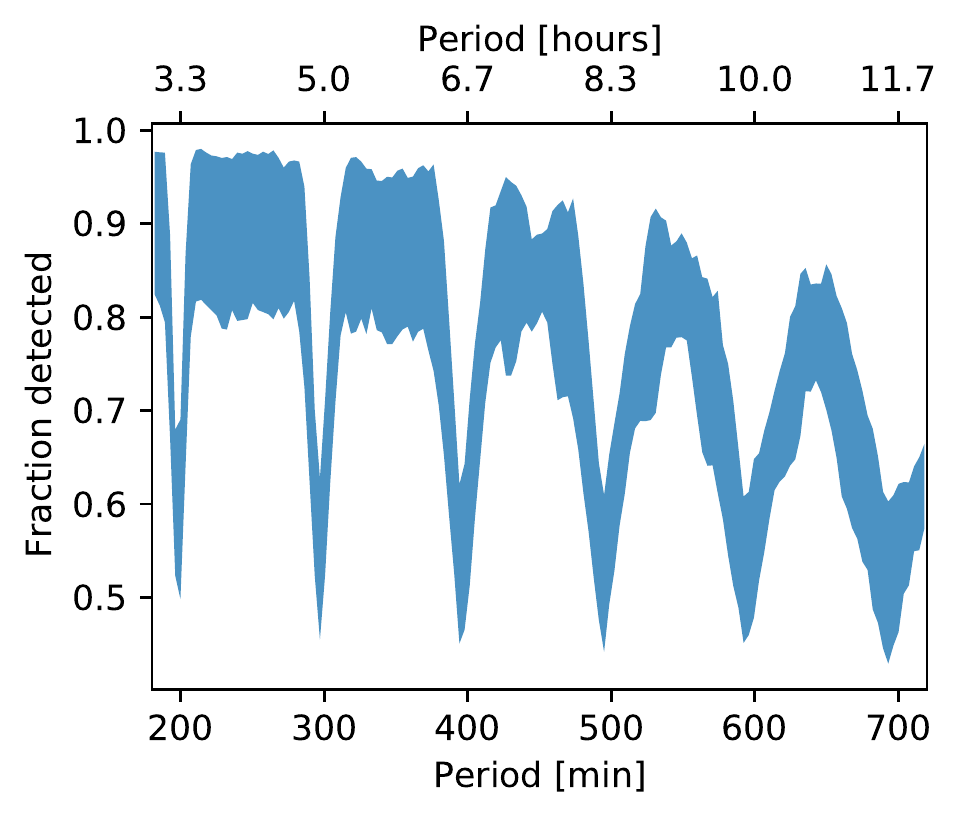}
    \caption{Fraction of detected, simulated and injected, transiting objects as a function of their orbital period. Aliases of the 100 minute orbital period of CHEOPS are visible as drops in detection efficiency near integer multiples of 100 minutes. The upper limit of the shaded region represents the fraction of objects detected for the brightest white dwarf in the sample, and the lower limit marks the detection fraction for the faintest white dwarf.}
    \label{fig:period}
\end{figure}

The period distribution of detected transiting objects is shown in Figure~\ref{fig:period}. Typical detection efficiency is near 90\% at orbital periods shorter than 5 hours, though it drops to 40-60\% when the orbital period of the transiting object is an integer multiple of the CHEOPS orbital period, and as the period approaches 12 hours. 

\begin{figure}
    \centering
    \includegraphics[scale=0.8]{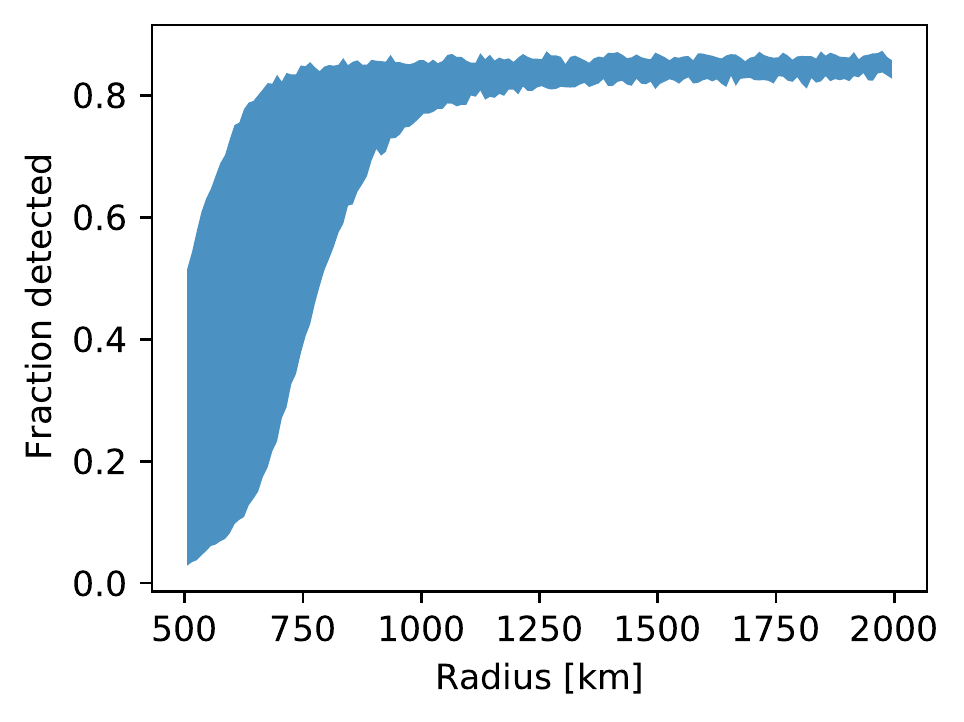}
    \caption{Fraction of detected, simulated and injected, transiting objects as a function of their radii. Objects with radii $>1000$ km are more likely than not to be detected with $S/N > 10$ around any of the six white dwarfs. As in Figure~\ref{fig:period}, the upper and lower limits of the shaded region mark the detected fraction of transiting objects for the brightest and faintest of the six white dwarfs. }
    \label{fig:radius}
\end{figure}

The radius distribution of detected objects is shown in Figure~\ref{fig:radius}. The BLS technique is most sensitive to objects roughly 1000 km in size or larger, about 60\% of the lunar radius. Objects of this size can be thought of as monolithic planetesimals, rather than small rubble-piles asteroids \citep{Lineweaver2010}.

\section{Discussion} \label{sec:discussion}

\subsection{DRP vs PIPE photometry}

We have presented results processed with the CHEOPS Data Reduction Pipeline (DRP) and with the custom PSF photometry pipeline \pipe, see Figures~\ref{fig:gallery} and \ref{fig:gallery_pipe} and the median absolute deviations (MADs) in each light curve in Table~\ref{tab:targets}. In most cases the DRP and \pipe photometry have similar MADs, though \pipe photometry yields smaller scatter than DRP photometry for the faintest targets by nearly a factor of two. We recommend PSF photometry for deep searches for transits with CHEOPS on faint targets ($V > 12$ mag), and for brighter targets the DRP photometry should suffice. 

\subsection{Transit probability}

The geometric transit probability for an Earth-sized planet orbiting a $0.6 M_\odot$ white dwarf in a 250 minute orbit is $3\%$, implying that $\sim30$ stars must be surveyed in order to have an appreciable chance of finding a single transiting object, assuming {\it all} white dwarfs host transiting material. As in \citet{Agol2011}, if we wanted to measure the white dwarf planet occurrence rate $\eta$ to a precision of $0.3$ (e.g., nine planets discovered total), we would need to survey $\sim 300 \eta^{-1}$ white dwarfs. Even in the optimistic scenario that $\eta\approx1$ for metal-polluted white dwarfs, we would require an observing program 50 times larger than the observations presented here to discover a new transiting planetesimal orbiting a white dwarf. For these reasons, we advocate here for continuing the transit search with wide-angle photometry from observatories like TESS \citep{Ricker2014}, LSST \citep{Lund2018}, or Evryscope \citep{Law2015}, which combined will measure hundreds of light curves of white dwarfs. 

Narrow, deep campaigns like the one presented here and in \cite{Wallach2018} for example, suggest that {\it if} transiting material were ubiquitous in orbit around white dwarfs, it would likely be quite small ($R_p \lesssim 1000$ km) or orbiting at periods longer than about $P \gtrsim$ 8 hours.

\subsection{Dust-free systems}

By adapting the survival models of \citet{Heng2010}, \citet{Veras2020} showed that transit detections are possible in systems without significant dust disks by delivery of intact planetesimals to close-in orbits at any cooling age. This reasoning suggests that transiting debris is perhaps likely orbiting white dwarfs regardless of metal pollution, so it may be fruitful to expand upon the work here to collect photometry of other bright, nearby white dwarfs that may host transiting debris without spectroscopic evidence of metal pollution.

\section{Conclusion} \label{sec:conclusion}

We present the first search for transiting planetismals orbiting white dwarfs with the high precision photometry mission CHEOPS. The spacecraft can perform sub-part-per-thousand photometry on targets as faint as $G = 12.75$ mag with minimal data processing steps by the user. We find that our small sample of six metal-polluted white dwarfs showed no significant transit-like events in the 24 hours that each target was observed. The exceptional quality photometry would have been sensitive to transiting material $\gtrsim1000$ km in radius on orbital periods $\lesssim5$ hours, but no such material was detected.

\begin{acknowledgements}
We are grateful for helpful mission support from Kate Isaak, and for valuable feedback from Jay Farihi.

CHEOPS is an ESA mission in partnership with Switzerland with important contributions to the payload and the ground segment from Austria, Belgium, France, Germany, Hungary, Italy, Portugal, Spain, Sweden, and the United Kingdom. 

This work has been carried out in the framework of the PlanetS National Centre of Competence in Research (NCCR) supported by the Swiss National Science Foundation (SNSF). This research has made use of the VizieR catalogue access tool, CDS, Strasbourg, France (DOI: 10.26093/cds/vizier). The original description of the VizieR service was published in A\&AS 143, 23. 

AS acknowledges support from a Science and Technology Facilities Council studentship.

We gratefully acknowledge the open source software which made this work possible: \textsf{astropy} \citep{Astropy2013, Astropy2018}, \textsf{ipython} \citep{ipython}, \textsf{numpy} \citep{numpy}, \textsf{scipy} \citep{scipy}, \textsf{matplotlib} \citep{matplotlib}, \textsf{synphot} \citep{synphot2018}, \textsf{batman} \citep{Kreidberg2015}, \textsf{lightkurve} \citep{lightkurve}

\end{acknowledgements}

\bibliographystyle{aa}
\bibliography{bibliography.bib}

\appendix

\section{Validation of the CHEOPS photometry with TESS} \label{app:a}

Within about a year of each of the CHEOPS observations of these white dwarfs, the TESS spacecraft observed four of the six white dwarfs studied in this work at two minute cadence \citep{Ricker2014}. We perform a parallel analysis of these TESS observations to demonstrate that the CHEOPS non-detections presented above are in fact confirmed by the lack of photometric variability in the TESS observations. 

\begin{figure*}
    \centering
    \includegraphics[scale=0.8]{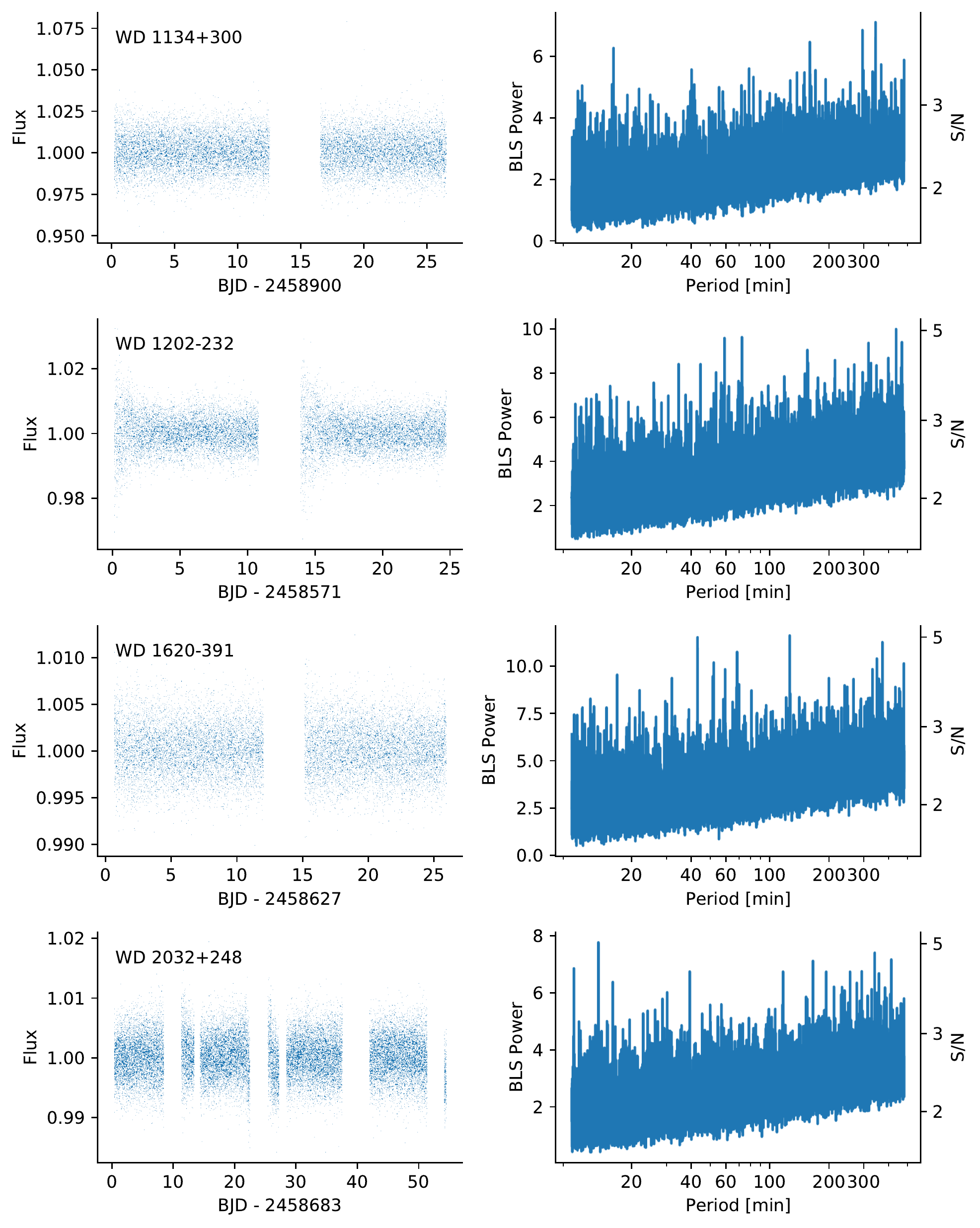}
    \caption{Same as Figure~\ref{fig:gallery} presenting the two-minute cadence TESS observations of four of the six white dwarfs. No significant $S/N > 5$ periodic events are detected, confirming the lack of variability reported with CHEOPS.}
    \label{fig:tess}
\end{figure*}

We perform a similar BLS periodogram analysis to the one presented in Section~\ref{sec:obs} on the TESS observations and plot the results in Figure~\ref{fig:tess}. No significant periodic events are detected with $S/N > 5$, indicating that indeed there is no transiting material orbiting these white dwarfs on periods of 10-300 minutes.

\section{Masked fluxes} \label{app:b}

\begin{figure*}
    \centering
    \includegraphics[scale=0.8]{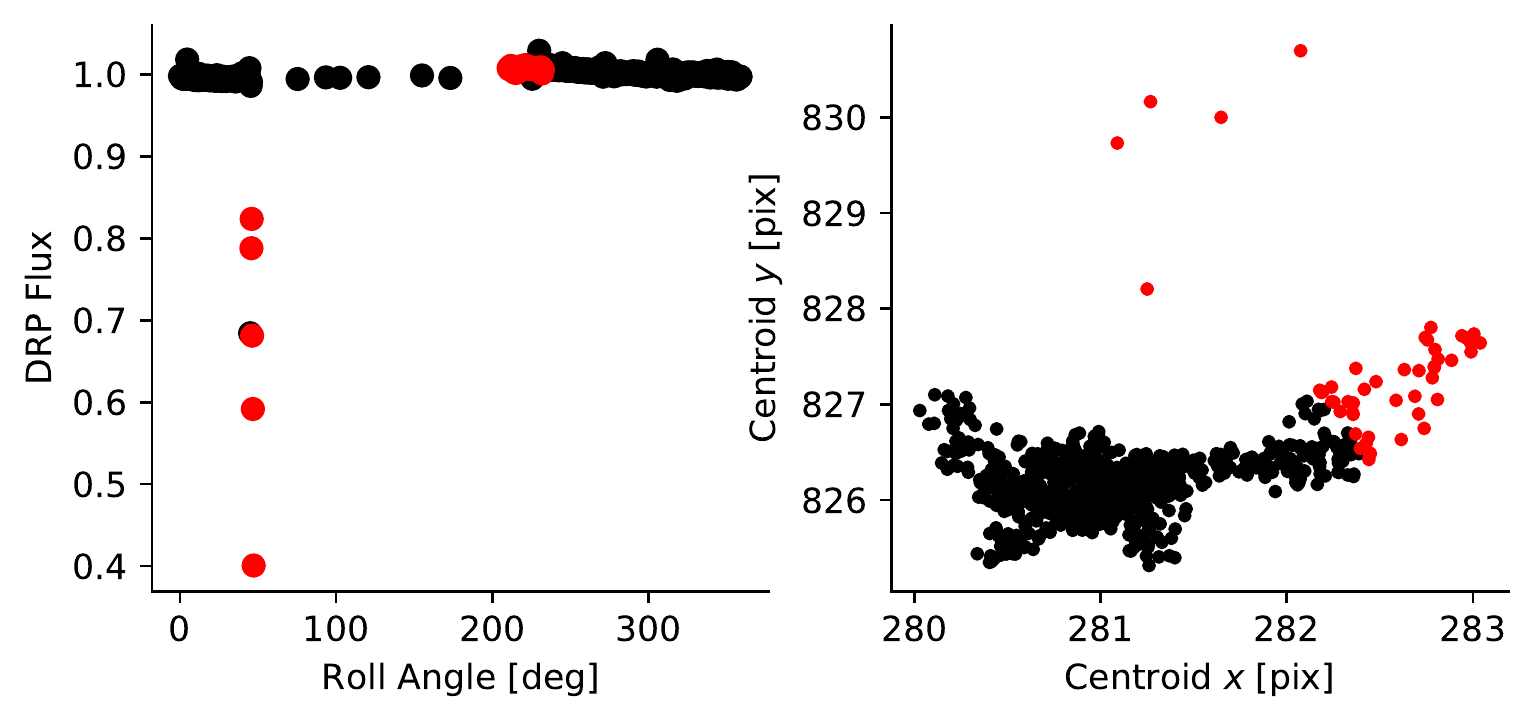}
    \caption{Masking of outlier fluxes (red) compared with typical observations (black) based on the $4\sigma$ radius about the mean centroid.}
    \label{fig:wd_roll}
\end{figure*}

Figure~\ref{fig:wd_roll} shows the flux masking applied to the observations of WD 0046+051. The most severe flux outliers occur when the stellar centroid is most distant from the mean centroid position, and also at the roll angle corresponding to the time immediately before Earth occultation near roll angle of $50^\circ$, where the expected scattered light is greatest.

\end{document}